\documentclass[%
 reprint,
 amsmath,amssymb,
 aps,
]{revtex4-2}

\usepackage{graphicx}% Include figure files
\usepackage{dcolumn}% Align table columns on decimal point
\usepackage{bm}% bold math
\usepackage{xcolor}
\usepackage{hyperref}% add hypertext capabilities

\begin{document}

\preprint{APS/123-QED}

\title{
%Adjoint master equation for out-of-time-ordered multi-time correlation functions
%\\
%Adjoint master equation for out-of-time-ordered correlators
Adjoint master equation for multi-time correlators
}

\author{Ivan V. Panyukov}
 \affiliation{Dukhov Research Institute of Automatics (VNIIA), 22 Sushchevskaya, Moscow 127055, Russia;}
 \affiliation{Moscow Institute of Physics and Technology, 9 Institutskiy pereulok, Dolgoprudny 141700, Moscow region, Russia;}
\author{Vladislav Yu. Shishkov}
 \email{vladislavmipt@gmail.com}
 \affiliation{Dukhov Research Institute of Automatics (VNIIA), 22 Sushchevskaya, Moscow 127055, Russia;}
 \affiliation{Moscow Institute of Physics and Technology, 9 Institutskiy pereulok, Dolgoprudny 141700, Moscow region, Russia;}
\author{Evgeny S. Andrianov}
 \affiliation{Dukhov Research Institute of Automatics (VNIIA), 22 Sushchevskaya, Moscow 127055, Russia;}
 \affiliation{Moscow Institute of Physics and Technology, 9 Institutskiy pereulok, Dolgoprudny 141700, Moscow region, Russia;}

\date{\today}% It is always \today, today,

\begin{abstract}
The quantum regression theorem is a powerful tool for calculating the muli-time correlators of operators of open quantum systems which dynamics can be described in Markovian approximation.
It enables to obtain the closed system of equation for the multi-time correlators.
%However, the possibility of obtaining such closed system of equation requires particular time ordering of the operators in multi-time correlators. 
However, the scope of the quantum regression theorem is limited by a particular time order of the operators in multi-time correlators and does not include out-of-time-ordered correlators. 
In this work, we obtain an adjoint master equation for multi-time correlators that is applicable to out-of-time-ordered correlators.
We show that this equation can be derived for various approaches to description of the dynamics of open quantum systems, such as the global or local approach. 
We show that the adjoint master equation for multi-time correlators is self-consistent.
Namely, the final equation does not depend on how the operators are grouped inside the correlator, and it coincides with the quantum regression theorem for the particular time ordering of the operators. 
\end{abstract}

\maketitle
 
\section{Introduction}
Out-of-time-ordered correlators (OTOCs) hold a special place in many-body quantum physics.
Originally, OTOC was introduced  in~\cite{larkin1969quasiclassical} to analyze the relation between the semi-classical and quantum description of the electrons in superconductors. 
In recent years, OTOCs receive growing attention because of their deep connection with the entanglement in quantum systems~\cite{shenker2014multiple, shenker2015stringy, luitz2017information, patel2017quantum, halpern2017jarzynski, syzranov2018out, garttner2018relating, nahum2018dynamics, lin2018out, garttner2018relating, khemani2018operator, rakovszky2018diffusive, halpern2018quasiprobability, roy2021entanglement}.
OTOCs can be used to quantitatively characterize the quantum chaos~\cite{roberts2015diagnosing, maldacena2016bound, maldacena2016remarks, stanford2016many, hashimoto2017out, chowdhury2017onset, tsuji2018out, chavez2019quantum, pilatowsky2020positive, fortes2020signatures, Garcia-Mata:2023} and to study some other problems in quantum systems~\cite{heyl2018detecting, de2019spectral, sun2020out, hu2021out, kirkova2022out, panyukov2022second, panyukov2023controlling}.
%Also, OTOCs determines correlation function of light emitted by quantum sources and transmitted through frequency filters~\cite{panyukov2022second, panyukov2023controlling}.
In addition, it is possible to measure OTOC experimentally~\cite{swingle2016measuring, zhu2016measurement, li2017measuring, garttner2017measuring, wei2018exploring, landsman2019verified, niknam2020sensitivity, sanchez2020perturbation, nie2020experimental, mi2021information, braumuller2022probing}. 

An experimentally feasible quantum system is never completely isolated from its environment. 
Therefore, the calculation of correlators should take into account the open nature of the quantum system. 
When the system dynamics can be described in Born--Markov approximation, the quantum regression theorem is applicable for calculations of multi-time correlation functions of operators.
However, the quantum regression theorem cannot be applied to OTOCs~\cite{breuer2002theory}.
In~\cite{blocher2019quantum}, the authors introduced the general quantum regression theorem (GQRT) for OTOC.  
The derivation of GQRT relies on Heisenberg-Langevin equations and requires explicit knowledge of the corresponding noise operators, which, in general, demands the knowledge of the systems eigenstates and eigenenergies.
Thus, it may be difficult to apply GQRT for complex quantum systems. 
%For complex quantum systems, the noise operators can be non-multiplicative~\cite{gardiner2004quantum} and eigenoperators~\cite{breuer2002theory} \textcolor{red}{why eigenoperators are necessary for GQRT from blocher2019?} can be unknown.
%In this case, .
%This derivation of GQRT relies on Heisenberg-Langevin equations.
%Thus, it requires explicit knowledge of the corresponding noise operators.
%For complex quantum systems, the noise operators can be non-multiplicative~\cite{gardiner2004quantum} and eigenoperators~\cite{breuer2002theory} \textcolor{red}{why eigenoperators are necessary for GQRT from blocher2019?} can be unknown.
%In this case, it is difficult to apply GQRT.

An alternative method to describe the dynamics of the open quantum system is the Lindblad master equation~\cite{breuer2002theory}.
In the Born--Markov approximation, the Heisenberg--Langevin equation and Lindblad master equation are equivalent~\cite{scully1997quantum}. 
Each of these two methods has its own advantages and disadvantages.
Usually, one of these methods is preferable for a particular problem.  
In the case, when the relaxation superoperators strongly depend on the interaction constant between subsystems of a complex open quantum system, it is necessary to investigate the ratios between the relaxation rates and this interaction constant.
When the interaction constant is smaller than relaxation rates, one uses so-called local approach~\cite{levy2014local, hofer2017markovian, gonzalez2017testing, shammah2018open, de2018reconciliation}, in the opposite case one uses global approach~\cite{breuer2002theory, levy2014local, hofer2017markovian, gonzalez2017testing, shishkov2019relaxation}.
In the intermediate case, one can use partially secular approach~\cite{cattaneo2019local, vovchenko2021model}, or the approach based on the perturbation theory~\cite{shishkov2020perturbation}. 
For this type of problem, the Lindblad master equation is preferable. 
In this regard, the important problem is to expand the scope of applicability of the quantum regression theorem to OTOCs in the formalism of Lindblad master equation.
%Note that quantum regression theorem in its standard from enables to obtain a closed system of equation for multi-time correlators from the Lindblad master equation.
%However, as has been noted above, the applicability of quantum regression theorem is restricted by the particular time ordering.

In this work, we derive the adjoint master equation for multi-time correlators.
We write explicitly the equation for OTOCs with an arbitrary number of operators.
We show that the adjoint master equation for multi-time correlators is self-consistent, such that the resultant equation does not depend on how we group the operators inside the trace function.
Derived adjoint master equation for multi-time correlators gives the same result as the standard quantum regression theorem when the latter is applicable.

\section{Decomposition of the system--reservoir interaction}
We consider the system interacting with the reservoir. 
The Hamiltonian of the system is $\hat H_S$, the Hamiltonian of the reservoir is $\hat H_R$, and the Hamiltonian of the system-reservoir interaction is $\hat H_{SR}$.
We assume that the evolution of the whole system is Hermitian and governed by the Hamiltonian
\begin{equation} \label{full Hamiltonian}
\hat H = \hat H_S + \hat H_R + \hat H_{SR}. 
\end{equation}
We consider the system Hamiltonian is time-independent. 
Additionally, we consider of interaction of the system with only one reservoir such that the interaction Hamiltonian can be presented in the form
\begin{equation} \label{SR Hamiltonian}
\hat H_{SR} = \hbar \lambda \hat S \hat R
\end{equation}
where $\hat S$ is the operator of the system, $\hat R$ is the operator of the reservoir, and $\lambda$ is the interaction constant.
One reservoir with the interaction Hamiltonian~(\ref{SR Hamiltonian}) is the simplest possible case, which, however, can be straightforwardly generalized to the multiple reservoirs. 

The Hamiltonian~(\ref{full Hamiltonian}) governs the time evolution of any operator $\hat D$ in Heisenberg representation
\begin{equation} \label{full evolution}
\hat D(t_2) = e^{i \hat H (t_2-t_1)/\hbar} \hat D(t_1) e^{-i \hat H (t_2-t_1)/\hbar}
\end{equation}
where $t_1$ and $t_2$ are arbitrary moments in time.
The equation for time evolution~(\ref{full evolution}) remains the same if the operator $\hat D$  is the system's operator, the reservoir's operator, or the product of both types.

In general case, it is rarely possible to establish an explicit form of the operator~(\ref{full evolution}), even if, at the initial moment, the operator $\hat D$ was only an operator of the system.

Usually, two conditions accompany the study of the open quantum system: (1) it is the evolution of the system that is important, whereas the evolution of the reservoir is not of interest; (2) the system has a small number of degrees of freedom, while the reservoir, on the contrary, has a large number of degrees of freedom.
When both conditions are met, one can exploit Born--Markov approximation and write effective equations for the averages and correlators for the system operators tracing out the reservoir's degrees.

To derive the equation for the correlators, we have to find the decomposition of the Herminian time evolution of the operator~$\hat S$ form the right-hand side of the Eq.~(\ref{SR Hamiltonian})~\cite{breuer2002theory}.
Typically, there are two possible options for such a decomposition.
The first one is the {\it exact} decomposition of the operator $\hat S$
\begin{equation} \label{exact decomposition}
e^{i \hat H_S t/\hbar} \hat S e^{-i \hat H_S t/\hbar} = \hat C_0 + \left( \sum_{j=1}^M \hat C_j e^{-i \omega_j t} + {h.c.}  \right).
\end{equation}
The second option is {\it approximate} decomposition of the operator $\hat S$
\begin{equation} \label{approximate decomposition}
e^{i \hat H_S t/\hbar} \hat S e^{-i \hat H_S t/\hbar} \approx \hat C_0 + \left( \sum_{j=1}^M \hat C_j e^{-i \omega_j t} + {h.c.}  \right).
\end{equation}
In Eq.~(\ref{exact decomposition})--(\ref{approximate decomposition}) $\omega_j \neq 0$, $\omega_j \neq \omega_k$ at $j \neq k$ and $\hat C_0^\dag = \hat C_0$.
The exact decomposition~(\ref{exact decomposition}) is used in the global approach to the LGKS equations~\cite{breuer2002theory, levy2014local, gonzalez2017testing, hofer2017markovian, shishkov2019relaxation}, whereas the approximate decomposition~(\ref{approximate decomposition}) is used in local approach~\cite{levy2014local, gonzalez2017testing, hofer2017markovian, shammah2018open, de2018reconciliation}, partially secular approach~\cite{vovchenko2021model}, and the approach based on perturbation theory~\cite{shishkov2020perturbation}.

\section{Interaction representation}
There are two coexisting parts that determine the time evolution of the correlators.
The first part originates from the Hamiltonian $\hat H_S$ and preserves the energy in the system. 
Thus, this part is associated with the Hermitian evolution. 
The second part originates from the Hamiltonians $\hat H_{SR}$ and $\hat H_R$, causing the energy flow between the system and reservoir or/and destruction of the phase in the system. 
Thus, this part is associated with non-Hermitian evolution.

We use the interaction representation to divide the Hermitian and non-Hermitian parts of the evolution. 
For arbitrary operator $\hat D$ its interaction representation $\hat D'(t)$ is defined by
\begin{equation} \label{interaction representation}
\hat D'(t)=e^{i(\hat H_S + \hat H_R)t/\hbar} \hat D e^{-i(\hat H_S + \hat H_R)t/\hbar}.
\end{equation}
The relation between the full quantum dynamics of the operator with its interaction representation follows from Eq.~(\ref{full evolution}) 
\begin{equation}  \label{interaction representation and full dynamics}
\hat D(t)= \hat V^\dag(t) \hat D'(t) \hat V(t).
\end{equation}
Here ${ \hat V(t) = e^{i (\hat H_S + \hat H_R) t/\hbar }e^{-i\hat H t/\hbar} }$ and it can be obtained from the equation 
\begin{equation} \label{evolution operator V}
{d \hat V(t) \over dt} =-{i \over \hbar} \hat H_{SR}'(t) \hat V(t)
\end{equation}
with the initial condition $\hat V(0) = \hat 1$.
According to Eq.~(\ref{SR Hamiltonian})~and~(\ref{interaction representation}), $\hat H^{\prime}_{SR}(t) = \hbar \lambda \hat S^{\prime}(t) \hat R^{\prime}(t)$.
Thus, we can explicitly find the approximate evolution of the operator $\hat V(t+\Delta t)$ up to second order in interaction constant  $\lambda$
\begin{equation} \label{approx V}
\hat V(t+\Delta t)
\approx
\hat V(t)
+
\hat W(t, \Delta t)
\end{equation}
where $W(t, \Delta t) = - i \lambda \hat V_1(t, \Delta t) -  \lambda^2 \hat V_2(t, \Delta t) $ and
\begin{equation} \label{V1 definition}
\hat V_1(t, \Delta t)= \int_{t}^{t+\Delta t}dt_1 \hat S'(t_1) \hat R'(t_1) \hat V(t), 
\end{equation}
\begin{multline} \label{V2 definition}
\hat V_2(t, \Delta t) = \\ 
\int_{t}^{t+\Delta t}dt_1\int_{t}^{t_1}dt_2~\hat S'(t_1) \hat S'(t_2) \hat R'(t_1) \hat R'(t_2) \hat V(t).
\end{multline}

The approximate time evolution~(\ref{approx V}) is the key to determining the effective non-Hermitian dynamics of the system.
Indeed, with this equation for any operator of the system $\hat B$, we can determine the relation between its values at time $t$ and at time $t+\Delta t$ 
\begin{multline} \label{main operator expansion}
\hat B(t + \Delta t) 
\approx 
\Delta t {i \over \hbar} \left[ \hat H_S(t), \hat B(t) \right] 
+
\\
 \left\{\hat V^\dag(t) + \hat W^\dag(t, \Delta t) \right\}
\hat B'(t)  
 \left\{\hat V(t) + \hat W(t, \Delta t) \right\}
.
\end{multline}
where we used $[\hat B(t), \hat H_R(t)] = 0$.
Note that we define the Hamiltonians $\hat H_S(t)$ and $\hat H_R(t)$ according to Eq.~(\ref{full evolution}).

We use Eq.~\ref{main operator expansion} to derive the quantum regression theorem and the more general adjoint master equation for multi-time correlators.

\section{Quantum regression theorem} \label{sec: QRT}
In this section, we derive the standard quantum regression theorem for the correlation function
\begin{equation} \label{1-time correlator}
\left\langle 
\hat A_1 \hat B_1(t+\tau) \hat A_2
\right\rangle
\end{equation}
where $\hat A_1$, $\hat B_1(t+\tau)$ and $\hat A_2$ are the operators of the system and $\hat A_1$ and $\hat A_2$ are taken at times prior to $t$ and do not depend on $\tau$.
We assume that we know the correlation finction~(\ref{1-time correlator}) at $\tau = 0$, and we are to determine its subsequent evolution at $\tau > 0$. 

To derive the quantum regression theorem, we consider the connection between the correlation functions $\langle \hat A_1 \hat B_1(t+\tau) \hat A_2 \rangle$ and $\langle \hat A_1 \hat B_1(t+\tau + \Delta \tau) \hat A_2 \rangle$.
Eq.~(\ref{main operator expansion}) leads to
\begin{multline} \label{1-time correlator expanded}
\left\langle 
\hat A_1 \hat B_1(t+\tau+\Delta\tau) \hat A_2
\right\rangle
\approx
\\
\Delta \tau {i \over \hbar} 
\left\langle 
\hat A_1 \left[\hat H_S(t+\tau), \hat B_1(t+\tau) \right] \hat A_2
\right\rangle
+
\\
\left\langle 
\hat A_1  \left\{\hat V^\dag(t + \tau) + \hat W^\dag(t + \tau, \Delta \tau)  \right\} \hat B'_1(t+\tau) 
\right.
\\
\left.
\left\{ \hat V(t + \tau) + \hat W(t + \tau, \Delta \tau) \right\} \hat A_2
\right\rangle
\end{multline}

For Eq.~(\ref{1-time correlator expanded}), we can apply assumptions and methods similar to those used for the derivation of the Lindblad master equation~\cite{breuer2002theory, shishkov2019relaxation}.
In particular, we set certain limitations for the time $\Delta \tau$, use the Born and Markov approximations, and assume that at any time $t$ the density matrix is factorized $\hat\rho = \hat\rho_S(t) \hat\rho_R^{\rm th}$, the operator of the reservoir has the zero mean ${\rm tr}_R(\hat R(t) \hat\rho_R^{\rm th}) = 0$.
We also suppose that we know either exact~(\ref{exact decomposition}) or approximate~(\ref{approximate decomposition}) decomposition of the system operator in interaction representation such that the operators $\hat C_j$ are known.
In the right-hand side of Eq.~(\ref{1-time correlator expanded}), we preserve only terms that either do not depend on $\Delta \tau$ or linear in $\Delta \tau$ and obtain the standard quantum regression theorem~(see Appendix~\ref{appendix: 1-time substitution})
\begin{equation} \label{standart QRT compact}
{
d\left\langle 
\hat A_1 \hat B_1(t+\tau) \hat A_2
\right\rangle
\over
d\tau
}
=
\left\langle 
\hat A_1 \mathcal{L}_{t+\tau}[\hat B_1(t+\tau)] \hat A_2
\right\rangle
\end{equation}
where we introduced the adjoint Lindblad superoperator
\begin{multline} \label{semigroup}
\mathcal{L}_{t}[\hat B] 
= 
{i \over \hbar} \left[\hat H_S(t), \hat B \right] + L_{\sqrt{\gamma_0} \hat C_0(t)} [\hat B] 
\\
+ 
\sum_{j=1}^M
L_{\sqrt{\gamma_j^\downarrow} \hat C_j(t)} [\hat B]
+
\sum_{j=1}^M
L_{\sqrt{\gamma_j^\uparrow} \hat C_j^\dag(t)} [\hat B]
\end{multline}
\begin{equation} \label{Lindblad superoperator}
L_{\hat C} [\hat B] 
= 
\hat C^\dag \hat B \hat C
-
{1 \over 2} \hat C^\dag \hat C \hat B
-
{1 \over 2} \hat B \hat C^\dag \hat C
\end{equation}
where the dissipation rates $\gamma_0$, $\gamma_j^\downarrow$ and $\gamma_j^\uparrow$ are considered in more details in Appendix~\ref{appendix: 1-time substitution}. 
%The dissipation rates $\gamma_j^\downarrow$ and $\gamma_j^\uparrow$ obey the Kubo--Martin--Schwinger relation for all $j$~\cite{breuer2002theory, kubo1957statistical, martin1959theory}.

%Consider some particular cases.
For $\hat A_1 = \hat 1$ and $\hat A_2 = \hat 1$, Eq.~(\ref{standart QRT compact}) reduces to the equation for the mean of the operator $\hat B_1(t+\tau)$.
For $\hat A_1 = \hat B^\dag(t)$, Eq.~(\ref{standart QRT compact}) leads to the standard quantum regression theorem for correlation functions $\langle \hat B_1^\dag (t) \hat B_1(t+\tau) \rangle$ that can be find in many textbooks~\cite{breuer2002theory, scully1997quantum}.

\section{Adjoint master equation for multi-time correlators}
The quantum regression theorem cannot be applied to calculate OTOCs. 
In this section, we adress this issue and derive adjoint master equation for multi-time correlators that is applicable to OTOCs.
The derivation itself is a generalization of the derivation of quantum regression theorem presented in the previous section.
To illustrate the main ideas behind this derivation we firstly consider a correlator with two time-dependent operators. 
Then we consider a correlator with arbitrary number of operators depended on time.

\subsection{Time evolution of 2 operators}
%The aim of this subsection is to highlight the difference in the derivation between the quantum regression theorem and the adjoint master equation for multi-time correlators.
%To do this, we consider the simplest correlator to which the quantum regression theorem is not applicable  
The simplest correlator to which the quantum regression theorem is not applicable, is 
\begin{equation} \label{2-time correlator}
\left\langle 
\hat A_1 \hat B_1(t+\tau) \hat A_2 \hat B_2(t+\tau) \hat A_3 
\right\rangle
\end{equation}
where $\hat A_1$, $\hat B_1(t+\tau)$, $\hat A_2$, $\hat B_2(t+\tau)$, and $\hat A_3$ are the operators of the system, and $\hat A_1$, $\hat A_2$, and $\hat A_3$ contain only times prior to $t$ and do not depend on $\tau$.
We assume we know the correlator~(\ref{2-time correlator}) at $\tau = 0$.

The connection between the correlator $$ \langle \hat A_1 \hat B_1(t+\tau) \hat A_2 \hat B_2(t+\tau) \hat A_3 \rangle $$ and the correlator $$ \langle \hat A_1 \hat B_1(t+\tau+\Delta\tau) \hat A_2 \hat B_2(t+\tau+\Delta\tau) \hat A_3 \rangle $$ follows from Eq.~(\ref{main operator expansion}) 
\begin{multline} \label{2-time correlator expanded}
\left\langle 
\hat A_1 \hat B_1(t+\tau+\Delta\tau) \hat A_2 \hat B_2(t+\tau+\Delta\tau) \hat A_3
\right\rangle
\approx
\\
\Delta \tau {i \over \hbar} 
\left\langle 
\hat A_1 \left[\hat H_S(t+\tau), \hat B_1(t+\tau)\right] \hat A_2 \hat B_2(t+\tau) \hat A_3
\right\rangle
+
\\
\Delta \tau {i \over \hbar} 
\left\langle 
\hat A_1 \hat B_1(t+\tau) \hat A_2 \left[\hat H_S(t+\tau), \hat B_2(t+\tau)\right] \hat A_3
\right\rangle
+
\\
\left\langle 
\hat A_1 
\left\{\hat V^\dag(t + \tau) + \hat W^\dag(t + \tau, \Delta \tau) \right\} 
\right.
\\
\left.
\hat B'_1(t+\tau) 
\left\{\hat V(t + \tau) + \hat W(t + \tau, \Delta \tau) \right\}
\right.
\\
\left.
\hat A_2 
\left\{\hat V^\dag(t + \tau) + \hat W^\dag(t + \tau, \Delta \tau) \right\}  
\right.
\\
\left.
\hat B'_2(t+\tau) 
\left\{\hat V(t + \tau) + \hat W(t + \tau, \Delta \tau) \right\} 
\hat A_3 
\right\rangle
\end{multline}
This equation allows us to obtain the adjoint master equation in almost the same way we derived the quantum regression theorem~(\ref{standart QRT compact}).
The details of this derivation are presented in Appendix~\ref{appendix: 2-time substitution}.
We note, that in this derivation we again suppose that we have either exact~(\ref{exact decomposition}) or approximate~(\ref{approximate decomposition}) decomposition such that the operators $\hat C_j$ are known.
There is one notable difference between this derivation and the derivation of the quantum regression theorem given in the previous section.
Namely, in this derivation, we have to take into account the combinations of operators $\hat W(t + \tau, \Delta \tau)$ and $\hat W^\dag(t + \tau, \Delta \tau)$ that belong to different operators $\hat B'_j(t+\tau)$ in the correlator~(\ref{2-time correlator expanded}). 
The resultant adjoint master equation for the correlator~(\ref{2-time correlator}) is
\begin{multline} \label{simple generalized QRT compact}
{
d\left\langle 
\hat A_1 \hat B_1(t+\tau) \hat A_2 \hat B_2(t+\tau) \hat A_3
\right\rangle
\over
d\tau
}
=
\\
\left\langle 
\hat A_1 \mathcal{L}_{t+\tau}[\hat B_1(t+\tau)] \hat A_2 \hat B_2(t+\tau) \hat A_3
\right\rangle
+
\\
\left\langle 
\hat A_1 \hat B_1(t+\tau) \hat A_2 \mathcal{L}_{t+\tau}[\hat B_2(t+\tau)] \hat A_3
\right\rangle
+
\\
\gamma_0
\left\langle 
\hat A_1 \left[\hat C_0, \hat B_1\right](t + \tau) 
\hat A_2 \left[\hat B_2, \hat C_0\right](t + \tau) \hat A_3 
\right\rangle
+
\\
\sum_{j=1}^M \gamma_j^\downarrow
\left\langle 
\hat A_1 \left[\hat C_j^\dag, \hat B_1\right](t + \tau) 
\hat A_2 \left[\hat B_2, \hat C_j\right](t + \tau) \hat A_3 
\right\rangle
+
\\
\sum_{j=1}^M \gamma_j^\uparrow
\left\langle 
\hat A_1 \left[\hat C_j, \hat B_1\right](t + \tau) 
\hat A_2 \left[\hat B_2, \hat C_j^\dag\right](t + \tau) \hat A_3 
\right\rangle
\end{multline}
where we denote ${ [\hat C, \hat B](t + \tau) = [\hat C(t + \tau), \hat B(t + \tau)] }$.
The last three terms in the right-hand side of the Eq.~(\ref{simple generalized QRT compact}) containing commutators between the operators $\hat C_j$ and $\hat B_m$ are the main difference between quantum regression and the adjoint master equation for multi-time correlators.

We stress, that in Eq.~(\ref{simple generalized QRT compact}) the operators $\hat C_j$ may correspond to the exact decomposition~(\ref{exact decomposition}) or to the approximate decomposition~(\ref{approximate decomposition}).
In both cases, the Eq.~(\ref{simple generalized QRT compact}) remains the same.

For $\hat A_1 = \hat 1$, Eq.~(\ref{simple generalized QRT compact}) reproduces the results obtained in~\cite{blocher2019quantum}.

\subsection{Time evolution of $n$ operators}
In this subsection, we present the adjoint master equation for the correlation function containing an arbitrary number of time-depended operators 
\begin{equation} \label{n-time correlator}
\left\langle 
\hat A_1 \hat B_1(t+\tau) ... \hat A_n \hat B_n(t+\tau) \hat A_{n+1} 
\right\rangle
\end{equation}
where $\hat A_j$, $\hat B_j(t+\tau)$ are the operators of the system and $\hat A_j$ may contain only times prior to $t$ and does not depend on $\tau$.
We assume that the correlation function~(\ref{n-time correlator}) at $\tau = 0$ is known.
Below, we also use the notation $ \left\langle \hat A_1 \hat B_1(t+\tau) ... \hat A_{n+1} \right\rangle $ for the correlation function~(\ref{n-time correlator}).

In this case, the derivation is the same as in the previous subsection.
As a result we obtain the adjoint master equation for the correlation function~(\ref{n-time correlator}) 
\begin{multline} \label{simple generalized QRT compact n-time}
{
d\left\langle 
\hat A_1 \hat B_1(t+\tau) ... \hat A_n \hat B_n(t+\tau) \hat A_{n+1} 
\right\rangle
\over
d\tau
}
=
\\
\sum_{m=1}^n
\mathcal{L}_{t+\tau}^{(j)}
\left[
\left\langle 
\hat A_1\hat B_1(t+\tau) ... \hat A_n \hat B_n(t+\tau) \hat A_{n+1} 
\right\rangle
\right]
+
\\
\sum_{m_1=1}^n
\sum_{m_2=m_1+1}^n
\mathcal{M}_{t+\tau}^{(m_1,m_2)}
\left[
\left\langle 
\hat A_1\hat B_1(t+\tau) ... \hat A_{n+1} 
\right\rangle
\right]
\end{multline}
where
\begin{multline}
\mathcal{L}_{t+\tau}^{(j)}
\left[
\left\langle 
\hat A_1\hat B_1(t+\tau) ... \hat A_n \hat B_n(t+\tau) \hat A_{n+1} 
\right\rangle
\right]
=
\\
\left\langle 
\hat A_1 ...  \hat A_j \mathcal{L}_{t+\tau}[\hat B_j(t+\tau)] \hat A_{j+1} ... \hat A_{n+1} 
\right\rangle
\end{multline}
\begin{multline}
\mathcal{M}_{t+\tau}^{(m_1,m_2)}
\left[
\left\langle 
\hat A_1\hat B_1(t+\tau) ... \hat A_{n+1} 
\right\rangle
\right]
=
\\
\gamma_0
\left\langle 
\hat A_1 ... [\hat C_0, \hat B_{m_1}](t+\tau) ... [\hat B_{m_2}, \hat C_0](t+\tau) ... \hat A_{n+1} 
\right\rangle
+
\\
\sum_{j=1}^M
\gamma_j^\downarrow
\left\langle 
\hat A_1 ... [\hat C_j^\dag, \hat B_{m_1}](t+\tau) ... [\hat B_{m_2}, \hat C_j](t+\tau) ... \hat A_{n+1} 
\right\rangle
+
\\
\sum_{j=1}^M
\gamma_j^\uparrow
\left\langle 
\hat A_1 ... [\hat C_j, \hat B_{m_1}](t+\tau) ... [\hat B_{m_2}, \hat C_j^\dag](t+\tau) ... \hat A_{n+1} 
\right\rangle
\end{multline}
The first term in the right-hand side of Eq.~(\ref{simple generalized QRT compact n-time}) corresponds to the dynamics of the solitary operators $\langle \hat B_j(t) \rangle$. 
The second term in the right-hand side of Eq.~(\ref{simple generalized QRT compact n-time}) describes the non-trivial time evolution of the correlator due to interplay between the operators $\hat B_j(t)$.
These terms are vital for OTOCs and provide the self-consistency of the adjoint master equation for multi-time correlators.

\section{Self-consistency of the adjoint master equation for multi-time correlators}
We show the self-consistency of the adjoint master equation for multi-time correlators for particular correlator
\begin{equation} \label{simple correlator for consistency}
\langle \hat B_1(t+\tau) \hat B_2(t+\tau) \rangle
\end{equation}
where $\hat B_1$ and $\hat B_2$ are the operators of the system. 

We can obtain the equation for this correlation function~(\ref{simple correlator for consistency}) in two different ways.
First, we can consider the correlator~(\ref{simple correlator for consistency}) as the correlator~(\ref{2-time correlator}) with $\hat A_1 = \hat A_2 = \hat A_3 = \hat 1$, $\hat B_2 = \hat 1$, and $\hat B_1 = (\hat B_1 \hat B_2)$.
In this case, Eq.~(\ref{simple generalized QRT compact}) gives 
\begin{equation} \label{version 1}
{
d \langle \hat B_1(t+\tau) \hat B_2(t+\tau) \rangle
\over
d \tau
}
=
\langle \mathcal{L}_{t+\tau}[\hat B_1(t+\tau) \hat B_2(t+\tau)] \rangle
\end{equation}
Second, we can consider the correlator~(\ref{simple correlator for consistency}) as the correlator~(\ref{2-time correlator}) with $\hat A_1 = \hat A_2 = \hat A_3 = \hat 1$. 
In this case, Eq.~(\ref{simple generalized QRT compact}) gives 
\begin{multline} \label{version 2}
{
d \langle \hat B_1(t+\tau) \hat B_2(t+\tau) \rangle
\over
d \tau
}
=
\\
\langle \mathcal{L}_{t+\tau}[\hat B_1(t+\tau)] \hat B_2(t+\tau) \rangle
+
\\
\langle \hat B_1(t+\tau) \mathcal{L}_{t+\tau}[\hat B_2(t+\tau)] \rangle
+
\\
\gamma_0
\left\langle 
[\hat C_0, \hat B_1](t + \tau) 
[\hat B_2, \hat C_0](t + \tau) 
\right\rangle
+
\\
\sum_{j=1}^M \gamma_j^\downarrow
\left\langle 
[\hat C_j^\dag, \hat B_1](t + \tau) 
[\hat B_2, \hat C_j](t + \tau)  
\right\rangle
+
\\
\sum_{j=1}^M \gamma_j^\uparrow
\left\langle 
[\hat C_j, \hat B_1](t + \tau) 
[\hat B_2, \hat C_j^\dag](t + \tau) 
\right\rangle
\end{multline}

The self-consistency of the adjoint master equation for multi-time correlators means that (1) Eq.~(\ref{version 1}) is equivalent to Eq.~(\ref{version 2}) and (2) adjoint master equation for multi-time correlators gives the same result as the standard quantum regression theorem~(\ref{standart QRT compact}) for the correlator~(\ref{1-time correlator}) with  $\hat A_1 = \hat A_2 = \hat 1$ and $\hat B_1 = (\hat B_1 \hat B_2)$.
One can easily prove both these statements.
Thus, the adjoint master equation for the correlation function~(\ref{simple correlator for consistency}) is self-consistent.

The proof of self-consistency of the adjoint master equation for more complex correlators is straightforward.

\section{Conclusion}
In this work, we derived adjoint master equation for multi-time correlators.
This equation is preferable for the usage of different approaches to relaxation operators such as global approach~\cite{breuer2002theory, shishkov2019relaxation, gonzalez2017testing, levy2014local, hofer2017markovian}, local approach~\cite{levy2014local, gonzalez2017testing, shammah2018open, hofer2017markovian, de2018reconciliation}, partially secular aproach~\cite{vovchenko2021model}, and the approach based on the perturbation theory~\cite{shishkov2020perturbation}. 
This is because the presented derivation does not depend whether exact or approximate expansion of the operator of interaction between system and reservoir is used.
Such flexibility reflects that the adjoint master equation for the multi-time correlators is independent of the particular relaxation operators.  

We showed that derived adjoint master equation for multi-time correlators is self-consistent: the final equation for a correlator does not depend on how we group the operators inside the correlator, and it coincides with the quantum regression theorem when the latter is applicable. 
This equation is not applicable to the out-of-time-ordered correlators in open quantum systems.
Also, adjoint master equation for multi-time correlators allows calculating of the correlator $\langle \hat A_1(t_1) \hat A_2(t_2) ...  \hat A_n(t_n) \rangle$ with arbitrary relations between $t_1$, $t_2$, ..., $t_n$.
The the exact algorithm in the case of $n=4$ is presented in~\cite{blocher2019quantum}.

In this work, we explicitly derived the adjoint master equation for one reservoir interacting with the system.
The derivation in the case of multiple reservoirs is straightforward.

\begin{acknowledgments}
The research was financially supported by a grant from Russian Science Foundation (project No. 20-72-10057).
V.Yu.Sh. thanks the Foundation for the Advancement of Theoretical Physics and Mathematics ``Basis''.
\end{acknowledgments}

\appendix

%\section{Derivation of Eq.~(\ref{main operator expansion})} \label{appendix: main expansion}
%Рассморим оператор системы $\hat B$, для которого $\hat B'(t) = \hat B e^{-i\omega_B t}$ и обозначим ${ \hat b(t) = e^{i\omega_B t} \hat B(t) }$.
%Тогда оператор $\hat b(t + \Delta t)$ и оператор $\hat b(t)$ связаны между собой согласно
%\begin{widetext}
%\begin{equation}
%\hat b(t + \Delta t) 
%\approx 
% \left\{\hat 1 + i\lambda \hat V_1^\dag(t, \Delta t) - \lambda^2 \hat V_2^\dag(t, \Delta t) \right\}
%\hat b(t)  
% \left\{\hat 1 - i\lambda \hat V_1(t, \Delta t) - \lambda^2 \hat V_2(t, \Delta t) \right\}
%\end{equation}
%\end{widetext}
%Отсюда легко получить уравнение~(\ref{main operator expansion}) с помощью тривиального обобщения на произвольный оператор системы $\hat B$.

\begin{widetext}
\section{Transformation of the correlators in the right-hand side of Eq.~(\ref{1-time correlator expanded}).} \label{appendix: 1-time substitution}
%Here, we present the transformation of the non-zero correlators depending on $\Delta \tau$ in the right-hand side of Eq.~(\ref{1-time correlator expanded}).
Here, we consider the second term in the right-hand side of Eq.~(\ref{1-time correlator expanded}),
\begin{equation} \label{term in 1-time correlator expanded}
\left\langle 
\hat A_1  \left\{\hat V^\dag(t + \tau) + \hat W^\dag(t + \tau, \Delta \tau)  \right\} \hat B'_1(t+\tau) 
\left\{ \hat V(t + \tau) + \hat W(t + \tau, \Delta \tau) \right\} \hat A_2
\right\rangle
\end{equation}
and trace out the reservoirs' degrees of freedom preserving the terms up to $\lambda^2$.
In this appendix, we use all the standard assumptions about the reservoir given in the textbook~\cite{breuer2002theory}. 
We also explicitly listed this assumptions in Section~\ref{sec: QRT}.
We separately consider the terms proportional to $\lambda^0$, $\lambda^1$ and $\lambda^2$.

To transform the term of~(\ref{term in 1-time correlator expanded}) proportional to $\lambda^0$ we use Eq.~(\ref{interaction representation and full dynamics}) and obtain
\begin{equation} \label{lambda0 in 1-time correlator expanded}
\left\langle 
\hat A_1  \hat V^\dag(t + \tau)  \hat B'_1(t+\tau) \hat V(t + \tau) \hat A_2
\right\rangle
=
\left\langle 
\hat A_1  \hat B_1(t+\tau) \hat A_2
\right\rangle
\end{equation}

All the terms of~(\ref{term in 1-time correlator expanded}) proportional to $\lambda^1$ are zero, because ${\rm tr}_R(\hat R(t) \hat\rho_R^{\rm th}) = 0$.
Namely,
\begin{equation} \label{lambda1_1 in 1-time correlator expanded}
\lambda^2
\left\langle 
\hat A_1 \hat V_1^\dag(t + \tau, \Delta \tau) \hat B'_1(t+\tau) \hat V(t + \tau) \hat A_2
\right\rangle
=
0
\end{equation}
\begin{equation} \label{lambda1_2 in 1-time correlator expanded}
\lambda^2
\left\langle 
\hat A_1 \hat V^\dag(t + \tau) \hat B'_1(t+\tau) \hat V_1(t + \tau, \Delta \tau)  \hat A_2
\right\rangle
=
0
\end{equation}

The terms of~(\ref{term in 1-time correlator expanded}) proportional to $\lambda^2$ are
\begin{equation} \label{corr_1}
\lambda^2
\left\langle 
\hat A_1 \hat V_1^\dag(t + \tau, \Delta \tau) \hat B'_1(t+\tau) \hat V_1(t + \tau, \Delta \tau) \hat A_2
\right\rangle
\end{equation}

\begin{equation} \label{corr_2}
\lambda^2
\left\langle 
\hat A_1 \hat V_2^\dag(t + \tau, \Delta \tau) \hat B'_1(t+\tau) \hat V(t + \tau) \hat A_2
\right\rangle
\end{equation}

\begin{equation} \label{corr_3}
\lambda^2
\left\langle 
\hat A_1 \hat V^\dag(t + \tau) \hat B'_1(t+\tau) \hat V_2(t + \tau, \Delta \tau) \hat A_2
\right\rangle
\end{equation}

We consider in details the correlator~Eq.~(\ref{corr_1}).
It can be transformed in the following way:
\begin{multline} \label{corr_1 details 1}
\lambda^2
\left\langle 
\hat A_1 \hat V_1^\dag(t + \tau, \Delta \tau) \hat B'_1(t+\tau) \hat V_1(t + \tau, \Delta \tau) \hat A_2
\right\rangle
=
\\
\lambda^2
\int_{t+\tau}^{t+\tau+\Delta \tau}dt_1 
\int_{t+\tau}^{t+\tau+\Delta \tau}dt_2 
\left\langle 
\hat A_1 
\hat V^\dag(t+\tau) \hat S'(t_1) \hat R'(t_1) 
\hat B'_1(t+\tau) 
\hat S'(t_2) \hat R'(t_2) \hat V(t+\tau) 
\hat A_2
\right\rangle
=
\\
\lambda^2
\int_{t+\tau}^{t+\tau+\Delta \tau}dt_1 
\int_{t+\tau}^{t+\tau+\Delta \tau}dt_2 
\left\langle 
\hat A_1 
\hat R'(t_1)
\left[
\hat C_0(t+\tau) + \sum_{j=1}^M \hat C_j(t+\tau) e^{-i \omega_j (t_1-t-\tau)} + \sum_{j=1}^M \hat C_j^\dag(t+\tau) e^{i \omega_j (t_1-t-\tau)} 
\right]
\right. \\ \left.
\hat B(t+\tau) 
\left[
\hat C_0(t+\tau) + \sum_{k=1}^M \hat C_k(t+\tau) e^{-i \omega_k (t_2-t-\tau)} + \sum_{k=1}^M \hat C_k^\dag(t+\tau) e^{i \omega_k (t_2-t-\tau)} 
\right]
\hat R'(t_2) 
\hat A_2
\right\rangle,
\end{multline}
where we used the definition Eq.~(\ref{V1 definition}), decomposition Eq.~(\ref{exact decomposition}) (or Eq.~(\ref{approximate decomposition})), and relation between interaction representation and full evolution given by Eq.~(\ref{interaction representation and full dynamics}).
We also replaced $\hat V^\dag(t+\tau)\hat R'(t_1)$ and $\hat R'(t_2)\hat V(t+\tau)$ with $\hat R'(t_1)\hat V^\dag(t+\tau)$ and $\hat V(t+\tau)\hat R'(t_2)$ for $t_1, t_2 \geq t$, because the corresponding commutators are proportional to $\lambda$ and can be omitted.
Following~\cite{breuer2002theory} we assume, that density matrix of the system and the reservoir are factorised at any moment in time. 
Thus, 
\begin{multline} \label{corr_1 details 2}
\lambda^2
\left\langle 
\hat A_1 \hat V_1^\dag(t + \tau, \Delta \tau) \hat B'_1(t+\tau) \hat V_1(t + \tau, \Delta \tau) \hat A_2
\right\rangle
=
\\
\lambda^2
\int_{0}^{\Delta \tau}d\tau_1 
\int_{0}^{\Delta \tau}d\tau_2 
\left\langle 
\hat R'(\tau_1)
\hat R'(\tau_2) 
\right\rangle
\left\langle 
\hat A_1 
\left[
\hat C_0(t+\tau) + \sum_{j=1}^M \hat C_j(t+\tau) e^{-i \omega_j \tau_1} + \sum_{j=1}^M \hat C_j^\dag(t+\tau) e^{i \omega_j \tau_1} 
\right]
\right. \\ \left.
\hat B(t+\tau) 
\left[
\hat C_0(t+\tau) + \sum_{k=1}^M \hat C_k(t+\tau) e^{-i \omega_k \tau_2} + \sum_{k=1}^M \hat C_k^\dag(t+\tau) e^{i \omega_k \tau_2} 
\right]
\hat A_2
\right\rangle,
\end{multline}
where we used $\left\langle \hat R'(\tau_1) \hat R'(\tau_2) \right\rangle = \left\langle \hat R'(t_1) \hat R'(t_2) \right\rangle$.
We integrate the correlator of the reservoir operators over time and obtain~\cite{breuer2002theory}
\begin{equation}
\lambda^2
\int_{0}^{\Delta \tau}d\tau_1 
\int_{0}^{\Delta \tau}d\tau_2 
\left\langle 
\hat R'(\tau_1)
\hat R'(\tau_2) 
\right\rangle
e^{ i \omega_A \tau_1}
e^{- i \omega_B \tau_2}
=
\Delta\tau \delta_{\omega_A, \omega_B}
\left\{
\begin{matrix}
\gamma_A^\downarrow, & \text{if } \omega_A > 0 \\
\gamma_A^\uparrow, & \text{if } \omega_A < 0 \\
\gamma_0, & \text{if } \omega_A = 0 
\end{matrix}
\right.
\end{equation}
Finally, we have
\begin{multline} \label{substitution 6}
\lambda^2
\left\langle 
\hat A_1 \hat V_1^\dag(t + \tau, \Delta \tau) \hat B'_1(t+\tau) \hat V_1(t + \tau, \Delta \tau) \hat A_2
\right\rangle
=
\Delta \tau
\gamma_0
\left\langle 
\hat A_1 \hat C_0(t + \tau) \hat B_1(t+\tau) \hat C_0(t + \tau) \hat A_2
\right\rangle
\\
+
\Delta \tau
\sum_{j=1}^M
\gamma_j^\downarrow
\left\langle 
\hat A_1 \hat C_j^\dag(t + \tau) \hat B_1(t+\tau) \hat C_j(t + \tau) \hat A_2
\right\rangle
+
\Delta \tau
\sum_{j=1}^M
\gamma_j^\uparrow
\left\langle 
\hat A_1 \hat C_j(t + \tau) \hat B_1(t+\tau) \hat C_j^\dag(t + \tau) \hat A_2
\right\rangle
\end{multline}

The consideration of the correlators~(\ref{corr_2}) and (\ref{corr_3}) is similar to (\ref{corr_1}).
Here, we present the result of the trasing out the reservoirs' degrees of freedom in these correlators 
\begin{multline} \label{substitution 7}
\lambda^2
\left\langle 
\hat A_1 \hat V_2^\dag(t + \tau, \Delta \tau) \hat B'_1(t+\tau) \hat V(t + \tau) \hat A_2
\right\rangle
=
\Delta \tau
{1 \over 2} 
\gamma_0
\left\langle 
\hat A_1 \hat C_0(t + \tau) \hat C_0(t + \tau) \hat B_1(t+\tau) \hat A_2
\right\rangle
+
\\
\Delta \tau
{1 \over 2} 
\sum_{j=1}^M
\gamma_j^\downarrow
\left\langle 
\hat A_1 \hat C_j^\dag(t + \tau) \hat C_j(t + \tau) \hat B_1(t+\tau) \hat A_2
\right\rangle
+
\Delta \tau
{1 \over 2} 
\sum_{j=1}^M
\gamma_j^\uparrow
\left\langle 
\hat A_1 \hat C_j(t + \tau) \hat C_j^\dag(t + \tau) \hat B_1(t+\tau) \hat A_2
\right\rangle
\end{multline}

\begin{multline} \label{substitution 8}
\lambda^2
\left\langle 
\hat A_1 \hat V^\dag(t + \tau) \hat B'_1(t+\tau) \hat V_2(t + \tau, \Delta \tau) \hat A_2
\right\rangle
=
\Delta \tau
{1 \over 2} 
\gamma_0
\left\langle 
\hat A_1 \hat B_1(t+\tau) \hat C_0(t + \tau) \hat C_0(t + \tau) \hat A_2
\right\rangle
+
\\
\Delta \tau
{1 \over 2} 
\sum_{j=1}^M
\gamma_j^\downarrow
\left\langle 
\hat A_1 \hat B_1(t+\tau) \hat C_j^\dag(t + \tau) \hat C_j(t + \tau) \hat A_2
\right\rangle
+
\Delta \tau
{1 \over 2} 
\sum_{j=0}^M
\gamma_j^\uparrow
\left\langle 
\hat A_1 \hat B_1(t+\tau) \hat C_j(t + \tau) \hat C_j^\dag(t + \tau) \hat A_2
\right\rangle
\end{multline}

Using the Eq.~(\ref{lambda0 in 1-time correlator expanded}), Eq.~(\ref{lambda1_1 in 1-time correlator expanded})--(\ref{lambda1_2 in 1-time correlator expanded}), and Eq.(\ref{substitution 6})--(\ref{substitution 8}), we derive the quantum regression theorem~(\ref{standart QRT compact}) from Eq.~(\ref{1-time correlator expanded}).

\section{Transformation of the correlators in the right-hand side of Eq.~(\ref{2-time correlator expanded}).} \label{appendix: 2-time substitution}
Here, we present the transformation of the correlator in the right-hand side of Eq.~(\ref{2-time correlator expanded}) that traces out the reservoirs' degrees of freedom and preserves the terms up to $\lambda^2$. 
We consider the correlator on the right-hand side of Eq.~(\ref{2-time correlator expanded}) that has the form
\begin{multline} \label{term in 2-time correlator expanded}
\left\langle 
\hat A_1 
\left\{\hat V^\dag(t + \tau) + \hat W^\dag(t + \tau, \Delta \tau) \right\} 
\hat B'_1(t+\tau) 
\left\{\hat V(t + \tau) + \hat W(t + \tau, \Delta \tau) \right\}
\hat A_2 
\right.  \\ \left.
\left\{\hat V^\dag(t + \tau) + \hat W^\dag(t + \tau, \Delta \tau) \right\}  
\hat B'_2(t+\tau) 
\left\{\hat V(t + \tau) + \hat W(t + \tau, \Delta \tau) \right\} 
\hat A_3 
\right\rangle
\end{multline}
and decompose it in correlators proportional to $\lambda^0$, $\lambda^1$, $\lambda^2$.

The term of the correlator~(\ref{term in 2-time correlator expanded}) proportional to $\lambda^0$ is
\begin{equation} \label{lambda0 term in 2-time correlator expanded}
\left\langle 
\hat A_1 
\hat V^\dag(t + \tau) 
\hat B'_1(t+\tau) 
\hat V(t + \tau)
\hat A_2 
\hat V^\dag(t + \tau)  
\hat B'_2(t+\tau) 
\hat V(t + \tau) 
\hat A_3 
\right\rangle
=
\left\langle 
\hat A_1 
\hat B_1(t+\tau) 
\hat A_2 
\hat B_2(t+\tau) 
\hat A_3 
\right\rangle
\end{equation}

All the terms of the correlator~(\ref{term in 2-time correlator expanded}) proportional to $\lambda^1$ are zero because ${\rm tr}_R(\hat R(t) \hat\rho_R^{\rm th}) = 0$.
Namely,
\begin{equation} \label{lambda1_1 term in 2-time correlator expanded}
\lambda
\left\langle 
\hat A_1 
\hat V_1^\dag(t + \tau, \Delta \tau) 
\hat B'_1(t+\tau) 
\hat V(t + \tau)
\hat A_2 
\hat V^\dag(t + \tau)  
\hat B'_2(t+\tau) 
\hat V(t + \tau) 
\hat A_3 
\right\rangle
=
0
\end{equation}

\begin{equation} \label{lambda1_2 term in 2-time correlator expanded}
\lambda
\left\langle 
\hat A_1 
\hat V^\dag(t + \tau)
\hat B'_1(t+\tau) 
\hat V_1(t + \tau, \Delta \tau) 
\hat A_2 
\hat V^\dag(t + \tau)  
\hat B'_2(t+\tau) 
\hat V(t + \tau) 
\hat A_3 
\right\rangle
=
0
\end{equation}

\begin{equation} \label{lambda1_3 term in 2-time correlator expanded}
\lambda
\left\langle 
\hat A_1 
\hat V^\dag(t + \tau)  
\hat B'_1(t+\tau) 
\hat V(t + \tau)
\hat A_2 
\hat V_1^\dag(t + \tau, \Delta \tau) 
\hat B'_2(t+\tau) 
\hat V(t + \tau) 
\hat A_3 
\right\rangle
=
0
\end{equation}

\begin{equation} \label{lambda1_4 term in 2-time correlator expanded}
\lambda
\left\langle 
\hat A_1 
\hat V^\dag(t + \tau) 
\hat B'_1(t+\tau) 
\hat V(t + \tau)
\hat A_2 
\hat V^\dag(t + \tau)  
\hat B'_2(t+\tau) 
\hat V_1(t + \tau, \Delta \tau) 
\hat A_3 
\right\rangle
=
0
\end{equation}

The terms of the correlator~(\ref{term in 2-time correlator expanded}) proportional to $\lambda^2$ are 
\begin{equation} \label{lambda2_1_1 term in 2-time correlator expanded}
\lambda^2
\left\langle 
\hat A_1 
\hat V_1^\dag(t + \tau, \Delta \tau) 
\hat B'_1(t+\tau) 
\hat V_1(t + \tau, \Delta \tau) 
\hat A_2 
\hat V^\dag(t + \tau)  
\hat B'_2(t+\tau) 
\hat V(t + \tau) 
\hat A_3 
\right\rangle
\end{equation}

\begin{equation} \label{lambda2_1_2 term in 2-time correlator expanded}
\lambda^2
\left\langle 
\hat A_1 
\hat V_2^\dag(t + \tau, \Delta \tau) 
\hat B'_1(t+\tau) 
\hat V(t + \tau) 
\hat A_2 
\hat V^\dag(t + \tau)  
\hat B'_2(t+\tau) 
\hat V(t + \tau) 
\hat A_3 
\right\rangle
\end{equation}

\begin{equation} \label{lambda2_1_3 term in 2-time correlator expanded}
\lambda^2
\left\langle 
\hat A_1 
\hat V^\dag(t + \tau) 
\hat B'_1(t+\tau) 
\hat V_2(t + \tau, \Delta \tau) 
\hat A_2 
\hat V^\dag(t + \tau)  
\hat B'_2(t+\tau) 
\hat V(t + \tau) 
\hat A_3 
\right\rangle
\end{equation}

\begin{equation} \label{lambda2_2_1 term in 2-time correlator expanded}
\lambda^2
\left\langle 
\hat A_1 
\hat V^\dag(t + \tau)  
\hat B'_1(t+\tau) 
\hat V(t + \tau) 
\hat A_2 
\hat V_1^\dag(t + \tau, \Delta \tau) 
\hat B'_2(t+\tau) 
\hat V_1(t + \tau, \Delta \tau) 
\hat A_3 
\right\rangle
\end{equation}

\begin{equation} \label{lambda2_2_2 term in 2-time correlator expanded}
\lambda^2
\left\langle 
\hat A_1 
\hat V^\dag(t + \tau)  
\hat B'_1(t+\tau) 
\hat V(t + \tau) 
\hat A_2 
\hat V_2^\dag(t + \tau, \Delta \tau) 
\hat B'_2(t+\tau) 
\hat V(t + \tau) 
\hat A_3 
\right\rangle
\end{equation}

\begin{equation} \label{lambda2_2_3 term in 2-time correlator expanded}
\lambda^2
\left\langle 
\hat A_1 
\hat V^\dag(t + \tau) 
\hat B'_1(t+\tau) 
\hat V(t + \tau) 
\hat A_2 
\hat V^\dag(t + \tau)  
\hat B'_2(t+\tau) 
\hat V_2(t + \tau, \Delta \tau) 
\hat A_3 
\right\rangle
\end{equation}

\begin{equation} \label{lambda2_3_1 term in 2-time correlator expanded}
\lambda^2
\left\langle 
\hat A_1 
\hat V_1^\dag(t + \tau, \Delta \tau) 
\hat B'_1(t+\tau) 
\hat V(t + \tau) 
\hat A_2 
\hat V_1^\dag(t + \tau, \Delta \tau) 
\hat B'_2(t+\tau) 
\hat V(t + \tau)  
\hat A_3 
\right\rangle
\end{equation}

\begin{equation} \label{lambda2_3_2 term in 2-time correlator expanded}
\lambda^2
\left\langle 
\hat A_1 
\hat V_1^\dag(t + \tau, \Delta \tau) 
\hat B'_1(t+\tau) 
\hat V(t + \tau) 
\hat A_2 
\hat V^\dag(t + \tau)  
\hat B'_2(t+\tau) 
\hat V_1(t + \tau, \Delta \tau) 
\hat A_3 
\right\rangle
\end{equation}

\begin{equation} \label{lambda2_3_3 term in 2-time correlator expanded}
\lambda^2
\left\langle 
\hat A_1 
\hat V^\dag(t + \tau)  
\hat B'_1(t+\tau) 
\hat V_1(t + \tau, \Delta \tau) 
\hat A_2 
\hat V_1^\dag(t + \tau, \Delta \tau) 
\hat B'_2(t+\tau) 
\hat A_3 
\hat V(t + \tau) 
\right\rangle
\end{equation}

\begin{equation} \label{lambda2_3_4 term in 2-time correlator expanded}
\lambda^2
\left\langle 
\hat A_1 
\hat V^\dag(t + \tau)  
\hat B'_1(t+\tau) 
\hat V_1(t + \tau, \Delta \tau) 
\hat A_2 
\hat V^\dag(t + \tau) 
\hat B'_2(t+\tau) 
\hat V_1(t + \tau, \Delta \tau) 
\hat A_3 
\right\rangle
\end{equation}

The correlators~(\ref{lambda2_1_1 term in 2-time correlator expanded})--(\ref{lambda2_1_3 term in 2-time correlator expanded}) are similar to the correlators (\ref{corr_1})--(\ref{corr_3}). 
These correlators correspond to the term 
\begin{equation}
\left\langle 
\hat A_1 \mathcal{L}_{t+\tau}[\hat B_1(t+\tau)] \hat A_2 \hat B_2(t+\tau) \hat A_3
\right\rangle
\end{equation}
in Eq.~(\ref{simple generalized QRT compact}).
The same is true for the correlators~(\ref{lambda2_2_1 term in 2-time correlator expanded})--(\ref{lambda2_2_3 term in 2-time correlator expanded}) that corrresponds to the term
\begin{equation}
\left\langle 
\hat A_1 \hat B_1(t+\tau) \hat A_2 \mathcal{L}_{t+\tau}[\hat B_2(t+\tau)] \hat A_3
\right\rangle
\end{equation}
in Eq.~(\ref{simple generalized QRT compact}).
The correlators (\ref{lambda2_3_1 term in 2-time correlator expanded})--(\ref{lambda2_3_4 term in 2-time correlator expanded}) do not directly correspond to the correlators (\ref{corr_1})--(\ref{corr_3}).
The consideration similar to one conducted for (\ref{corr_1}) in Appendix~\ref{appendix: 1-time substitution} can be done for the correlators (\ref{lambda2_3_1 term in 2-time correlator expanded})--(\ref{lambda2_3_4 term in 2-time correlator expanded}) leading to the following result
\begin{multline}
\lambda^2
\left\langle 
\hat A_1 
\hat V_1^\dag(t + \tau, \Delta \tau) 
\hat B'_1(t+\tau) 
\hat V(t + \tau) 
\hat A_2 
\hat V_1^\dag(t + \tau, \Delta \tau) 
\hat B'_2(t+\tau) 
\hat V(t + \tau)  
\hat A_3 
\right\rangle
=
\\
\Delta \tau
\gamma_0
\left\langle 
\hat A_1 \hat C_0(t + \tau) \hat B_1(t+\tau) 
\hat A_2 \hat C_0(t + \tau)  \hat B_2(t+\tau) \hat A_3 
\right\rangle
\\
+
\Delta \tau
\sum_{j=1}^M \gamma_j^\downarrow
\left\langle 
\hat A_1 \hat C_j^\dag(t + \tau) \hat B_1(t+\tau) 
\hat A_2 \hat C_j(t + \tau)  \hat B_2(t+\tau) \hat A_3 
\right\rangle
\\
+
\Delta \tau
\sum_{j=1}^M \gamma_j^\uparrow
\left\langle 
\hat A_1 \hat C_j(t + \tau) \hat B_1(t+\tau) 
\hat A_2 \hat C_j^\dag(t + \tau)  \hat B_2(t+\tau) \hat A_3 
\right\rangle
\end{multline}

\begin{multline}
\lambda^2
\left\langle 
\hat A_1 
\hat V_1^\dag(t + \tau, \Delta \tau) 
\hat B'_1(t+\tau) 
\hat V(t + \tau) 
\hat A_2 
\hat V^\dag(t + \tau)  
\hat B'_2(t+\tau) 
\hat V_1(t + \tau, \Delta \tau) 
\hat A_3 
\right\rangle
=
\\
\Delta \tau
\gamma_0
\left\langle 
\hat A_1 \hat C_0(t + \tau) \hat B_1(t+\tau) 
\hat A_2 \hat B_2(t+\tau) \hat C_0(t + \tau) \hat A_3 
\right\rangle
\\
+
\Delta \tau
\sum_{j=1}^M \gamma_j^\downarrow
\left\langle 
\hat A_1 \hat C_j^\dag(t + \tau) \hat B_1(t+\tau) 
\hat A_2 \hat B_2(t+\tau) \hat C_j(t + \tau) \hat A_3 
\right\rangle
\\
+
\Delta \tau
\sum_{j=1}^M \gamma_j^\uparrow
\left\langle 
\hat A_1 \hat C_j(t + \tau) \hat B_1(t+\tau) 
\hat A_2 \hat B_2(t+\tau) \hat C_j^\dag(t + \tau) \hat A_3 
\right\rangle
\end{multline}

\begin{multline}
\lambda^2
\left\langle 
\hat A_1 
\hat V^\dag(t + \tau)  
\hat B'_1(t+\tau) 
\hat V_1(t + \tau, \Delta \tau) 
\hat A_2 
\hat V_1^\dag(t + \tau, \Delta \tau) 
\hat B'_2(t+\tau) 
\hat A_3 
\hat V(t + \tau) 
\right\rangle
=
\\
\Delta \tau
\gamma_0
\left\langle 
\hat A_1 \hat B_1(t+\tau) \hat C_0(t + \tau) 
\hat A_2 \hat C_0(t + \tau) \hat B_2(t+\tau) \hat A_3 
\right\rangle
\\
+
\Delta \tau
\sum_{j=1}^M \gamma_j^\downarrow
\left\langle 
\hat A_1 \hat B_1(t+\tau) \hat C_j^\dag(t + \tau) 
\hat A_2 \hat C_j(t + \tau) \hat B_2(t+\tau) \hat A_3 
\right\rangle
\\
+
\Delta \tau
\sum_{j=1}^M \gamma_j^\uparrow
\left\langle 
\hat A_1 \hat B_1(t+\tau) \hat C_j(t + \tau) 
\hat A_2 \hat C_j^\dag(t + \tau) \hat B_2(t+\tau) \hat A_3 
\right\rangle
\end{multline}

\begin{multline}
\lambda^2
\left\langle 
\hat A_1 
\hat V^\dag(t + \tau)  
\hat B'_1(t+\tau) 
\hat V_1(t + \tau, \Delta \tau) 
\hat A_2 
\hat V^\dag(t + \tau) 
\hat B'_2(t+\tau) 
\hat V_1(t + \tau, \Delta \tau) 
\hat A_3 
\right\rangle
=
\\
\Delta \tau
\gamma_0
\left\langle 
\hat A_1 \hat B_1(t+\tau) \hat C_0(t + \tau) 
\hat A_2 \hat B_2(t+\tau) \hat C_0(t + \tau) \hat A_3 
\right\rangle
\\
+
\Delta \tau
\sum_{j=1}^M \gamma_j^\downarrow
\left\langle 
\hat A_1 \hat B_1(t+\tau) \hat C_j^\dag(t + \tau) 
\hat A_2 \hat B_2(t+\tau) \hat C_j(t + \tau) \hat A_3 
\right\rangle
\\
+
\Delta \tau
\sum_{j=1}^M \gamma_j^\uparrow
\left\langle 
\hat A_1 \hat B_1(t+\tau) \hat C_j(t + \tau) 
\hat A_2 \hat B_2(t+\tau) \hat C_j^\dag(t + \tau) \hat A_3 
\right\rangle
\end{multline}
These correlators correspond to the terms 
\begin{multline}
\gamma_0
\left\langle 
\hat A_1 \left[\hat C_0, \hat B_1\right](t + \tau) 
\hat A_2 \left[\hat B_2, \hat C_0\right](t + \tau) \hat A_3 
\right\rangle
+
\sum_{j=1}^M \gamma_j^\downarrow
\left\langle 
\hat A_1 \left[\hat C_j^\dag, \hat B_1\right](t + \tau) 
\hat A_2 \left[\hat B_2, \hat C_j\right](t + \tau) \hat A_3 
\right\rangle
+
\\
\sum_{j=1}^M \gamma_j^\uparrow
\left\langle 
\hat A_1 \left[\hat C_j, \hat B_1\right](t + \tau) 
\hat A_2 \left[\hat B_2, \hat C_j^\dag\right](t + \tau) \hat A_3 
\right\rangle
\end{multline}
in Eq.~(\ref{simple generalized QRT compact}).

\end{widetext}

\bibliography{QRT}

\end{document}